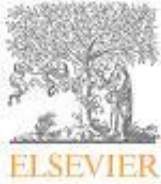

# Development of a magnet power supply with a common-mode cancellation method based on automatic calibration of driving pulses

Yonggao Zhang [a], Chengkai Chai [a], Peng Liu [a] [*], Fengjun Wu [b]

*[a]East China Jiaotong University, Nanchang 330013, China*

*[b]Institue of Modern Physics, Lanzhou 730000, China*

**Abstract**

Based on conventional H-bridge topologies of magnet power supplies, a high-frequency common-mode equivalent model was established. It is pointed out that the two phase legs have complementary effect of common-mode noise under bipolar PWM modulation. Due to the phase difference of the driving pulses in the real product, however, the effect of common-mode cancellation is greatly disrupted. Based on the common-mode equivalent model, a frequency domain analysis was carried out to determine the theoretical relationship between the common-mode noise intensity and the phase difference, which can guide the adjustment of the driving pulses. It was also found that the increase of switching frequency will aggravate the common-mode noise caused by the phase difference. To reduce the common-mode noise, a method that automatic calibration of the driving pulses was evaluated by a simulation and experimentally. Finally, a prototype is built, and the results of the prototype show that the proposed method can effectively reduce the common-mode noise.

## 1. Introduction

With the development of a new generation of switching devices, wide bandgap devices such as silicon carbide and gallium nitride have been gradually applied to the magnet power supply [1]. Higher switching frequency improves the performance of magnet power supplies, however, it also causes serious electromagnetic interference (EMI) to the system. The EMI consists of mainly common-mode and differential-mode noises [2]. The influence factors of common-mode noise are more complex, and the suppression of its high frequency band is more difficult. Because of common-mode noise, the efficiency of the system will be reduced, and the normal-mode excitation current will also be disturbed. This will affect the orbit of the beam in the accelerator [3]. One method to suppress common-mode noise is passive or active filtering [4-5]. A neutral point clamped(NPC) inverter with reduced switching sequences is given in [6]. Compared with conventional H-bridge topologies, the NPC topology could sufficiently reduce the common-mode noise current.

---

[*] Corresponding author. e-mail: liupen9@ustc.edu.cn (P. Liu).

In this paper, a common-mode equivalent model with parasitic parameters of H-bridge circuits, which are widely used in accelerator magnet power supplies, is established, and the reason of common-mode current is analyzed. It is found that no common-mode current is induced in principle by the bipolar PWM modulation strategy. In practice, however, the phase difference at the switching time of the power switch devices will be produced due to the dispersion of the driving modules, which could cause the time-varying common-mode voltage at high frequency and generate the common-mode noise current. So based on the high-frequency common-mode equivalent model, the effects of phase difference and switching frequency on the spectrum characteristics of common-mode current are discussed, and the effective value of common-mode current after rectification and filtering is used as feedback information to adjust the driving pulses, so as to suppress the common-mode current. Finally, the method is validated by experiment, which shows that the common-mode noise current can be reduced substantially.

## 2. Common-mode noise

Magnet power supplies usually consist of two parts: rectifier and chopper. To simplify the analysis, the rectifier part is equivalent to an ideal DC voltage source. The equivalent circuit model of magnet power supplies with parasitic capacitance is shown in Fig. 1, where the $U_{dc}$ is the input voltage of chopper, and $U_{dc}=E$. $C_{in}$ is the electrolytic capacitor at the input side of the H-bridge circuit, $L_1$ and $L_2$ are the filter inductors, $C_o$ is the filter capacitor, $Z_m$ is the equivalent load (magnet inductance and magnet resistance). $S_1$-$S_4$ are semi-conductor switches such as IGBTs. P and N represent the positive and negative output ports of the H-bridge circuit, respectively. $C_{d1}$ and $C_{d2}$ are the equivalent parasitic capacitances to ground between the output side of front stage rectifier and the electrolytic capacitor. Due to the internal structure of IGBTs and the interconnection between IGBTs, heat sinks and the grounding case, there are parasitic capacitors $C_{S1}$ and $C_{S2}$ of switches to ground at P and N terminals respectively. $C_{C1}$ and $C_{C2}$ are the parasitic cable capacitances [5]. According to the definition of common-mode voltage $u_{CM}$ and differential-mode voltage $u_{DM}$, the negative electrode B of $U_{dc}$ is taken as the reference point, which can be expressed by Eq. (1) and Eq. (2).

$$u_{CM}=\frac{u_{PB}+u_{NB}}{2} \quad (1)$$

$$u_{DM}=u_{PB}-u_{NB} \quad (2)$$

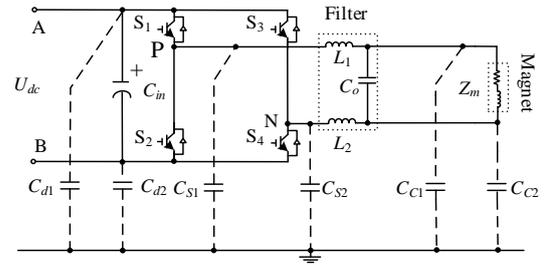

Fig. 1. Circuit of magnet power supplies with stray parameters.

Using common-mode voltage and differential-mode voltage to represent the voltage of P and N terminals, and substituting them into Fig. 1, the common-mode equivalent circuit of Fig. 1 can be obtained as shown in Fig. 2a in theory of circuit superposition principle, in which the related devices of the differential-mode loop have been eliminated. Fig. 2b shows the simplified model of Fig. 2a. Eqs. (3), (4), (5) and (6) show that $Z_{D1}$, $Z_{D2}$, $u_{D1}$, $u_{D2}$.

$$Z_{D1}=\frac{\alpha\beta}{\alpha+\beta}+\frac{Z_{CC1}Z_{CC2}}{Z_m+Z_{CC1}+Z_{CC2}} \quad (3)$$

$$Z_{D2}=\frac{Z_{CS1}Z_{CS2}}{Z_{CS1}+Z_{CS2}} \quad (4)$$

$$u_{D1}=\frac{u_{DM}}{2}\cdot\frac{\beta-\alpha}{\alpha+\beta} \quad (5)$$

$$u_{D2}=\frac{u_{DM}}{2}\frac{Z_{CS2}-Z_{CS1}}{Z_{CS2}+Z_{CS1}} \quad (6)$$

Where the $\alpha$, $\beta$ are determined by the expressions

$$\alpha = Z_{L1} + \frac{Z_m Z_{CC1}}{Z_m + Z_{CC1} + Z_{CC2}} \quad (7)$$

$$\beta = Z_{L2} + \frac{Z_m Z_{CC2}}{Z_m + Z_{CC1} + Z_{CC2}} \quad (8)$$

In order to further intuitively characterize the influencing factors of common-mode current, the most simplified model can be obtained by the electric circuit equivalent transformation of Fig. 2b, as shown in Fig. 2c. The equivalent differential-mode voltage $u'_{DM}$ and the equivalent common-mode impedance $Z$ are shown in Eqs. (9), (10).

$$u'_{DM} = \frac{u_{D1} Z_{D2} + u_{D2} Z_{D1}}{Z_{D1} + Z_{D2}}$$
$$= \frac{u_{DM}}{2(Z_{D1} + Z_{D2})} \left[ \frac{(\beta - \alpha) Z_{D2}}{\alpha + \beta} + \frac{(Z_{CS2} - Z_{CS1}) Z_{D1}}{Z_{CS1} + Z_{CS2}} \right] \quad (9)$$

$$Z = \frac{Z_{D2} Z_{D1}}{Z_{D1} + Z_{D2}} + Z_{CD} \quad (10)$$

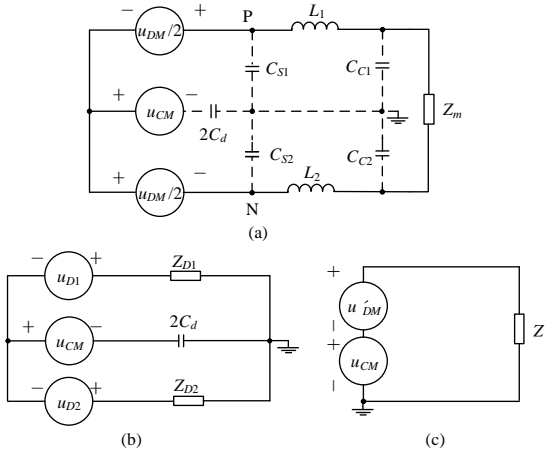

Fig. 2. Common-mode circuit model for magnet power supplies.

It can be concluded that the common-mode current can be equivalent to the common-mode impedance $Z$ generated by the common-mode voltage $u_{CM}$ and the equivalent differential-mode voltage $u'_{DM}$ by the above analysis, where $u'_{DM}$ is determined by the differential-mode voltage $u_{DM}$, the circuit components and parasitic parameters. It is not difficult to draw the following conclusions by the formula (7) - (9).

(1) When the system is symmetrically arranged and the parasitic parameters are strictly consistent ($Z_{CS1} = Z_{CS2}$, $Z_{CC1} = Z_{CC2}$), the amplitude of the equivalent differential-mode voltage is 0. So the common-mode current can be eliminated by keeping $u_{CM}$ constant.

(2) If $u_{CM}$ is a time-varying voltage at high-frequency, it is necessary to adjust the circuit structure and its related parameters to keep the sum of $u_{CM}$ and $u'_{DM}$ constant to eliminate the common-mode current.

Fig. 3 shows the PWM control signals and the midpoint voltages of phase legs when the bipolar modulation is used. It can be seen that, in ideal case, the switch state of the switch tubes in the diagonal position is the same, the two switch tubes in the same bridge arm are complementary, and the sum of $u_{PB}$ and $u_{NB}$ is constant $E$, that is, $u_{CM}$ is always constant. Therefore, the bipolar PWM modulation strategy can eliminate the common-mode current.

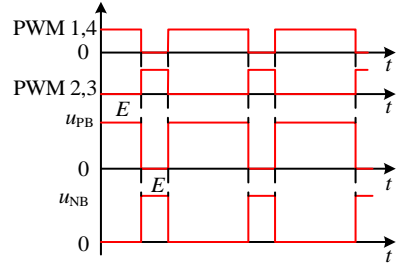

Fig. 3. PWM control signals and midpoint voltages of phase legs

But in a real product, the PWM pulses have to go through the driving circuit to control the power devices. The dispersion of these signal conditioning circuits will lead to hundreds of nanoseconds phase difference between the driving signals of the two power devices on the diagonal position of the H-bridge circuit. Then the operational sequences will be increased from two to four in each switch cycle. The voltages of the circuit for each operational sequence are shown in Table 1.

Table 1

Voltages of the circuit with different PWM phases.

| operational sequence | voltages of the circuit | | | |
|---|---|---|---|---|
| | $u_{PB}$ | $u_{NB}$ | $u_{CM}$ | $u_{DM}$ |
| $T_1$ | $E$ | $E$ | $E$ | 0 |
| $T_2$ | $E$ | 0 | $E/2$ | $E$ |
| $T_3$ | 0 | 0 | 0 | 0 |
| $T_4$ | 0 | $E$ | $E/2$ | $-E$ |

Table 1 shows that no matter how to adjust the circuit parameters, the sum of $u_{CM}$ and $u'_{DM}$ cannot be kept as constant, because it is always not equal in the operational sequences of $T_1$ and $T_3$. Therefore, the phase inconsistency of the driving pulses will affect the common-mode elimination effect of the bipolar PWM modulation.

## 3. Common-mode cancellation method by automatic calibration of driving pulses

The premise of common-mode rejection by automatic calibration of driving pulses is to determine the relationship between the common-mode noise level and the phase difference of driving pulses. The spectrum characteristics of common-mode current can characterize the interference intensity in detail, so this section will study the influence of phase difference time on the spectrum characteristics of common-mode current based on the established high-frequency common-mode equivalent model.

Assuming that the relevant parasitic parameters are consistent, then by substituting the relevant parameters of the main circuit in Table 2 into Eqs. (3), (4), (7), (8), and (10), we can obtain the operational impedance $Z(s)$ of the common-mode equivalent circuit in Fig.2c, as approximately shown in Eq. (11).

$$Z(s)=(25.1s^4 +8\times10^5 s^3 +1\times10^{14} s^2 +1.2\times10^{18} s \\ +3.2\times10^{21})/(5\times10^4 s^3 +5.4\times10^8 s^2 +1.5\times10^{12} s) \quad (11)$$

The $Z(s)$ Bode diagram is shown in Fig. 4. It can be seen that the operational impedance $Z(s)$ is the minimum at the frequency of $3.34\times10^5$Hz.

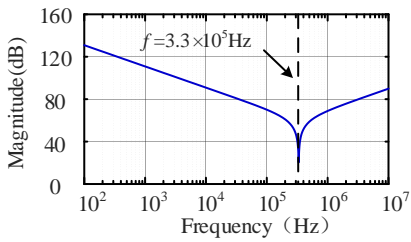

Fig. 4. Bode diagram of $Z(s)$.

Further, The waveform of common-mode voltage is drawn through Table 1 when the driving pulses are inconsistent, as shown in Fig. 5. And there are

$$t_1 = \frac{(1-d)T-|\tau|}{2} \quad (12)$$

$$t_2 = \frac{(1-d)T+|\tau|}{2} \quad (13)$$

Where T is the switching period, $d$ is the duty ratio and $\tau$ is the phase difference (the positive and negative of $\tau$ represent the lag/lead of driving pulses respectively).

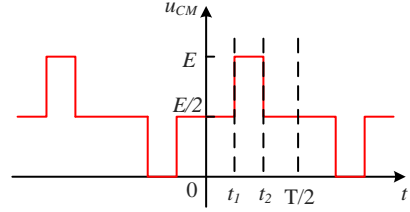

Fig. 5. Waveform of common-mode voltage

The Fourier transform of $u_{CM}(t)$ is shown in Eq. (14), where $\omega_0 = \frac{2\pi}{T}$.

$$u_{CM}(t) = \frac{E}{2} + \sum_{k=1}^{\infty} \frac{2E}{k\pi} \sin(k\pi - k\pi d)\sin\left(k\pi \times \frac{|\tau|}{T}\right)\sin(k\omega_0 t) \quad (14)$$

Relevant parasitic parameters are symmetric and consistent, so the total excitation voltage $u(t)$ of the common-mode interference is

$$u(t)=u_{CM}(t)+u'_{DM}(t) = u_{CM}(t) \quad (15)$$

The $k$th harmonic amplitude of common-mode current $I_{CM}^{(k)}(j\omega)$ can be obtained by eliminating the DC component in $u(t)$,

$$I_{CM}^{(k)}(j\omega) = \frac{\left|\frac{2E}{k\pi}\sin(k\pi - k\pi d)\sin\left(k\pi \times \frac{|\tau|}{T}\right)\right|}{\left|Z^{(k)}(j\omega)\right|} \quad (16)$$

It can be seen from Eq. (16) that when the system impedance network remains unchanged, the spectrum amplitude of common-mode current is not only proportional to the DC voltage of H-bridge, E, but also related to the switching period T, duty ratio $d$ and phase difference $\tau$. Fig. 6- Fig. 8 show the spectrum characteristic of common-mode current when the switching frequency ($f_s$) is 10kHz, 20kHz and 40kHz with different duty ratios (where the harmonic order is based on the switching frequency).

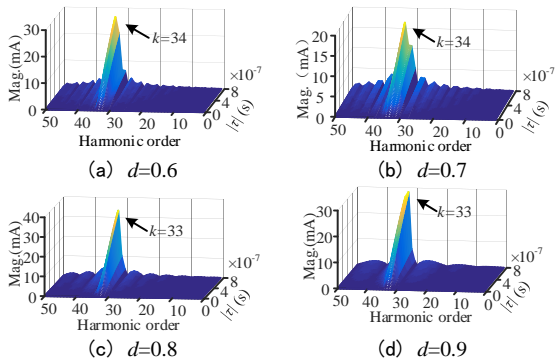

Fig. 6. Spectrum of common-mode current with different duty ratios ($f_S$=10kHz)

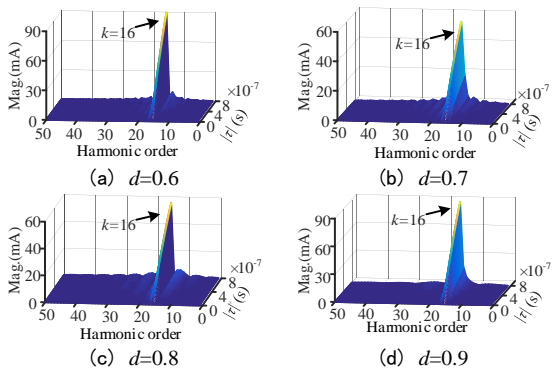

Fig. 7. Spectrum of common-mode current with different duty ratios ($f_S$=20kHz)

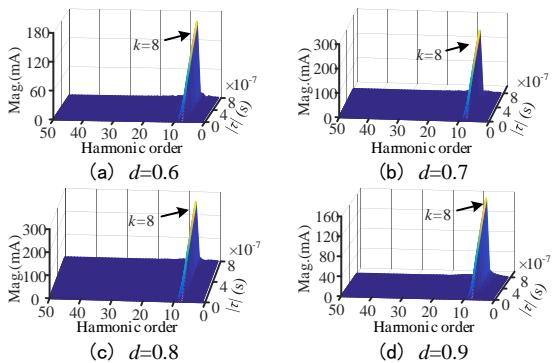

Fig. 8. Spectrum of common-mode current with different duty ratios ($f_S$=40kHz)

It can be seen from the three groups of figures that the spectrum amplitude of common-mode current varies with different duty ratios under the same switching frequency. The peak frequency of common-mode current spectrum is near the frequency point where the total impedance of system equivalent network is the minimum, and the common-mode conducted interference is mainly concentrated in the frequency range of 200kHz - 500kHz. Comparing the spectrum characteristics of common-mode current with different switching frequencies under the same duty ratio, it can be found that the main harmonic amplitude of common-mode current increases by times with the increase of switching frequency, and the common-mode electromagnetic interference of the system is more serious. Therefore, the switching frequency of the circuit should be considered comprehensively and should not be too high.

And the amplitude of each harmonic of the common-mode current increases significantly with the increase of $|\tau|$ (the phase difference) of the driving signals under the same switching frequency. Theoretically, when $|\tau|=0$, the amplitude of each harmonic is zero. Therefore, this paper proposes a method of adjusting the driving pulses, which tries to make the phase difference of the pulses zero, so as to suppressing the common-mode current.

For the actual products of different magnet power supplies, the pulses phase difference of power switch devices must be different. Therefore, in order to improve the portability of phase adjustment technology and the real-time adjustment of the driving pulses phase, the closed-loop control method is preferred. Fig. 9 is the phase adjustment control diagram.

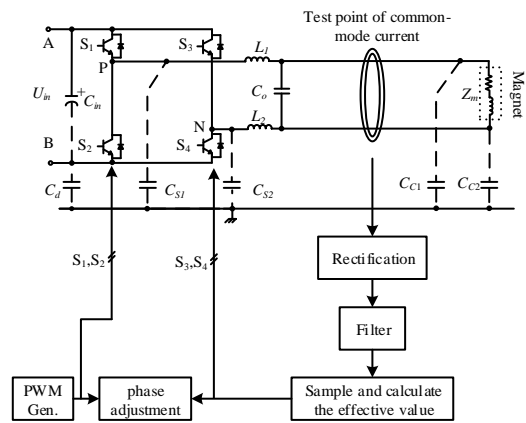

Fig. 9 Phase adjustment of driving pulses

A detailed analysis of the spectrum of common-mode current requires a complex calculation process, which is a very heavy task for DSP. Therefore, it is a good solution to calculate the effective value of the signal from the perspective of time domain (Eq. (17)) and use it as feedback information.

$$I_{CM} = \sqrt{\frac{1}{N}\sum_{n=0}^{N-1}|i_{CM}[n]|^2} \quad (17)$$

The oscillation frequency of common-mode current can reach up to several MHz, which is difficult to detect directly and effectively by DSP. Therefore, this paper designs a rectifier and filter circuit to convert high frequency common-mode signal to DC signal and then send it to DSP for acquisition and calculation, which can reduce the workload of DSP. Next, taking the common-mode current at a certain frequency as an example, it shows that the conditioned DC signal can reflect the common-mode interference intensity effectively.

Let the common-mode current component at this frequency be

$$i_{CMn} = A\sin\omega_n t \quad (18)$$

Then the effective value is

$$I_{CMn} = \frac{\sqrt{2}}{2}A \quad (19)$$

The waveform function after rectification is as follows

$$i'_{CMn} = A|\sin\omega_n t| = \frac{2}{\pi}A + \frac{4A}{\pi} \times \left(\frac{1}{1\times3}\cos 2\omega_n t - \frac{1}{3\times5}\cos 4\omega_n t + \frac{1}{5\times7}\cos 6\omega_n t....\right) \quad (20)$$

The filtered signal is the DC component of $i'_{CMn}$, so the effective value of the final signal sent to DSP is

$$I'_{CMn} = \frac{2}{\pi}A = \frac{2\sqrt{2}}{\pi}I_{CMn} \quad (21)$$

It can be seen from Eq. (21) that after the common-mode current is rectified and filtered, its effective value is directly proportional to the effective value of the original signal, so the numerical change of the effective value of the DC signal can reflect the strength change of the common-mode interference.

From the foregoing analysis, it can be seen that the frequency spectrum amplitude of common-mode current increases monotonically with $|\tau|$ increasing, and there is a minimum value when $|\tau|=0$. Therefore, the problem of common-mode current suppression can be regarded as the problem of finding the optimal solution of the model. And the system is optimal when $|\tau|=0$.

So the basic idea of suppressing common-mode interference is to first give an initial value (time adjustment) in any direction (lead / lag), and then take the effective value of DC signal as feedback information to judge whether the common-mode current decreases. If it is favorable, increase a step in this direction; If not, adjust a step in the opposite direction. After several adjustments, the common-mode current can be effectively suppressed. The flow chart is shown in Fig 10.

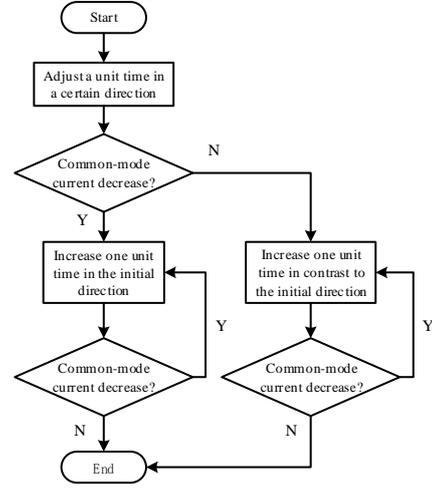

Fig. 10 Flow chart of the common-mode current suppression

## 4. System evaluation

In order to verify the correctness of the theoretical analysis of the common-mode equivalent model and the effectiveness of the common-mode current suppression method, an experimental prototype is built as shown in Fig. 11. The main parameters are shown in Table 2.

The DSP is a TMS320F28335. Based on the two switches $S_1$ and $S_2$, the driving pulses of the switches $S_3$ and $S_4$ of the right bridge arm are adjusted, so that the switches on the diagonal position can be turned on and off at the same time. Due to the limitation of laboratory conditions, the load of the magnet is simulated by series connection of a resistance and an inductance.

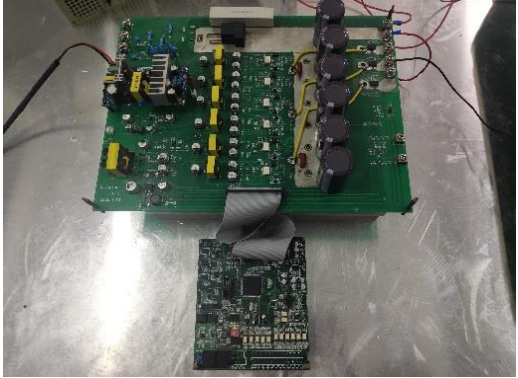

Fig. 11 Prototype of the magnet power supply

Table 2
Parameters of the experimental prototype.

| | |
|---|---|
| Filter inductance | 2mH |
| Filter capacitance | 1mF |
| Equivalent inductance of the magnet | 5mH |
| Equivalent resistance of the magnet | 25Ω |
| IGBT | IRG7PH42UD1 |
| Charging voltage | 150V |
| Output current | 4A |
| Switching frequency | 10kHz |

Fig. 12 shows the comparison of gate driving voltage waveforms ($u_{GE1}$, $u_{GE4}$) and collector voltage waveforms ($u_{CE1}$, $u_{CE4}$) of switches $S_1$ and $S_4$ before and after adjustment. It can be seen from the figure that in the actual device, although the phase of the PWM pulse sent by the DSP is equal, there is a time difference of about 380ns for the driving signal sent to the IGBT gate, which also causes the switching action of $S_4$ to lag 400ns compared with $S_1$. However, after pre tuning the PWM signal of DSP, there is no phase difference between the two switching devices.

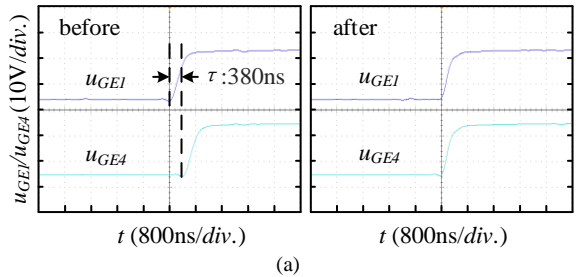

(a)

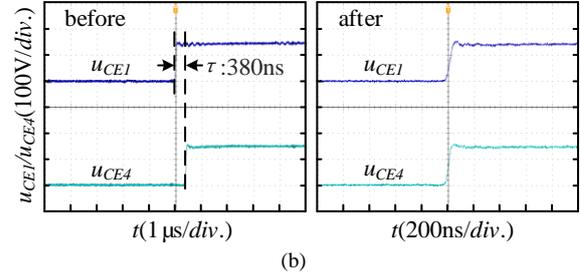

(b)

Fig. 12 $u_{GE}$ (a) and $u_{CE}$ (b) of IGBT before and after tuning

Fig. 13 shows the voltage ($u_{PB}$, $u_{NB}$) between the output midpoint of the two bridge arms and the negative pole of the DC bus, the common-mode voltage ($2u_{CM}$) and the local enlarged charts before and after adjustment. It can be seen that the switching time of the power devices at the diagonal position is inconsistent, which leads to obvious common-mode voltage spike with the amplitude of 150V (the actual common-mode voltage amplitude is 75V), and this spike width is the lag time of $S_4$ relative to $S_1$, which is basically consistent with the common-mode voltage waveform shown in Fig. 5. After tuning the phase of driving pulses, the common-mode voltage keeps stable and the peak voltage is greatly suppressed.

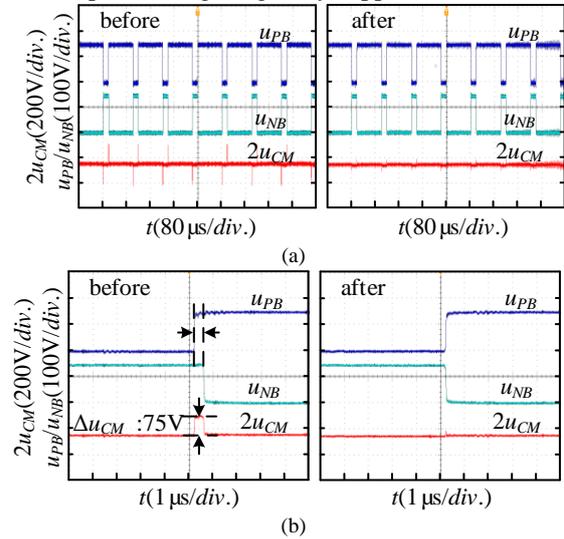

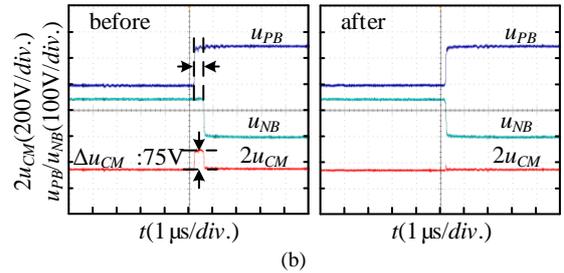

Fig. 13 $u_{PB}$, $u_{NB}$, $u_{CM}$ of IGBT (a) and local enlarged charts (b) before and after tuning

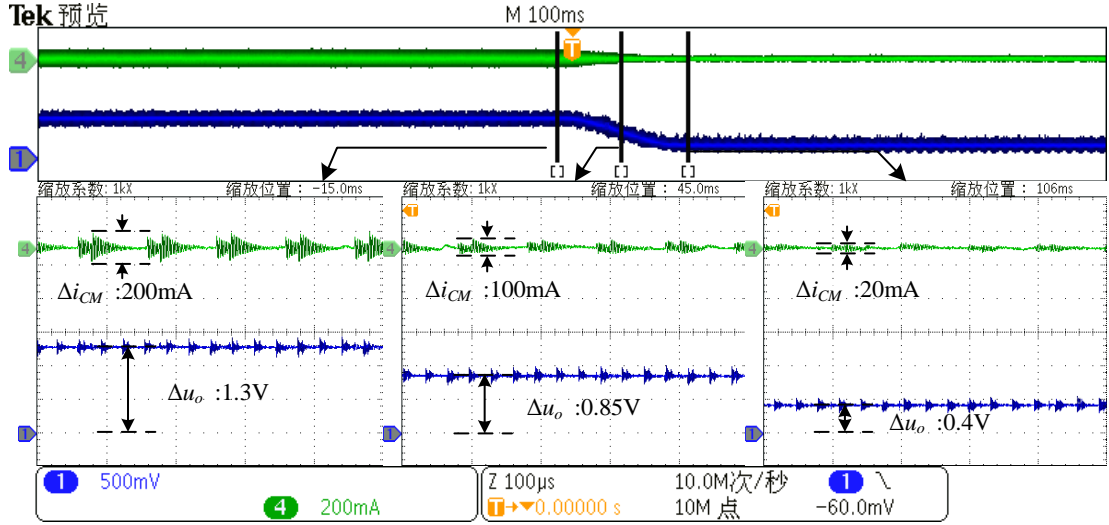

Fig. 14. Waveform of common mode current in the attenuation process

In order to verify the effectiveness of pulse phase closed-loop adjustment and the correctness of the common -mode current rectifier and filter circuit, the attenuation process of common-mode current with program operation is tested. The local enlarged waveforms of common-mode current (channel 4, green) and rectifier filter signal (channel 1, blue) at different stages are shown in Fig. 14.

It can be seen that the amplitude of DC signal after rectification and filtering will gradually decrease in the process of common-mode current attenuation. Before common- mode suppression, the peak to peak value of common mode current is 200mA, and the DC signal amplitude is 1.3V after rectification and filtering; In attenuation process 1, the peak to peak value of common-mode current decreases to 120mA, and the amplitude of corresponding signal after rectification and filtering is 1.0V; In attenuation process 2, the common-mode current is further reduced to 100mA, and the DC signal amplitude is reduced to 0.85V; After the common-mode suppression program is completed, the common-mode current is reduced to 40mA, and the DC signal amplitude reaches the lowest 0.4V. The reason why the common-mode current is not completely eliminated here is that the adjustment step of DSP is limited, and the parasitic parameters of the actual circuit are inconsistent, which makes the differential-mode voltage produce a certain degree of common-mode current. But in general, after the phase adjustment of closed-loop control, the common-mode current is effectively suppressed. The results also show the correctness of the common -mode current rectifier and filter circuit.

The common-mode current, its spectrum and output current waveform of the system are shown in Fig. 15. It can be seen from the figure that there is a common-mode current with peak to peak value of 200mA in the actual system. However, after the closed-loop adjustment by the DSP, the common-mode current decreases obviously. From the spectrum characteristics of common-mode current, the harmonic amplitude of each main frequency is also greatly reduced.

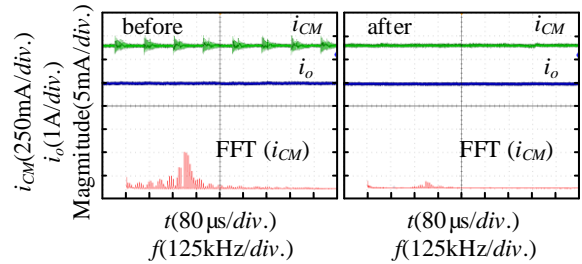

Fig. 15. $i_o$, $i_{CM}$ and its spectrum before and after tuning($f_S$=10kHz)

In order to compare the effectiveness of the closed-loop control system with the common-mode current rejection at different switching frequencies, the measured waveforms of the circuit at 20kHz are shown in Fig. 16.

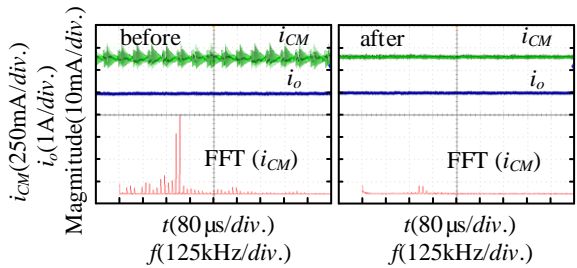

Fig. 16. $i_o$, $i_{CM}$ and its spectrum before and after tuning($f_s$=20kHz)

As can be seen from Fig. 16, after tuning the phase of driving pulses, the peak to peak value of the common-mode current decreases from 275mA to 43mA. The peak value of spectrum is reduced from 35mA to 4mA, and the harmonic amplitude of each frequency is greatly reduced.

Fig.17 shows the measured waveforms of the circuit at 40kHz. From Fig.17, we can see that when the switching frequency of 40kHz is adopted, the common-mode interference of the system is more serious, and the peak to peak value of common-mode current reaches 398mA. However, after the closed-loop adjustment of the driving pulses, the common-mode current is reduced to 75mA, which is 81% lower than before. The peak value of common-mode current spectrum is also reduced from 46 mA to 8 mA, and the suppression effect is good.

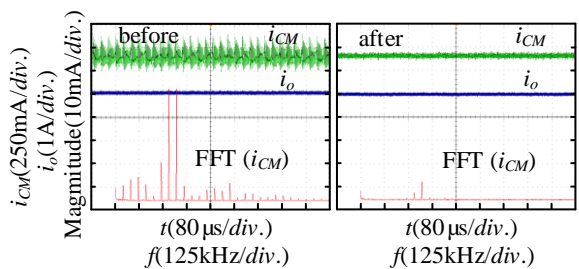

Fig. 17. $i_o$, $i_{CM}$ and its spectrum before and after tuning($f_s$=40kHz)

## 5. Conclusion

In this paper, a high-frequency common-mode equivalent model is established by analyzing the interference source and the common-mode current transmission path of the H-bridge topology of DC magnet power supplies. According to the model, the spectrum of common-mode current is analyzed. The analysis shows that the phase delay of driving pulses will greatly affect the effect of common-mode cancellation in bipolar modulation. From the spectrum characteristics, the harmonic amplitude of common-mode current is positively correlated with the phase delay time. And the rise of switching frequency will aggravate the common-mode interference caused by the phase delay. For the problem that the high frequency common-mode current is difficult to collect, this paper proposes the idea of rectifier and filtering to process it, and takes the effective value of the signal as the evaluation index of the common-mode interference intensity to determine the phase control time of the driving signals. Finally, experiments show the effectiveness of the method. And the proposed method could sufficiently reduce the common-mode noise of the magnet power supply at different switching frequencies.